\documentclass[%
reprint,
aps,
prb,
floatfix,
]{revtex4-2}

\usepackage{amsmath}
\usepackage{amssymb}
\usepackage{mathtools} 

\usepackage{siunitx}
\usepackage{bm}
\usepackage[version=4]{mhchem}

\usepackage[usenames,dvipsnames]{xcolor}
\usepackage{ulem} 
\usepackage{comment}

\usepackage{graphicx}
\usepackage{epsfig}
\usepackage{subfigure}

\usepackage{array}
\usepackage{longtable}
\usepackage{tabularx}
\usepackage{booktabs}

\usepackage{hyperref}
\usepackage{pdfpages}
\makeatletter
\AtBeginDocument{\let\LS@rot\@undefined}
\makeatother

\begin{document}

\title{CrRhAs: a member of a large family of metallic kagome antiferromagnets}

\author{Y. N. Huang}
\affiliation{Department of Physics, Zhejiang University of Science and Technology, Hangzhou 310023, People’s Republic of China}

\author{Harald O. Jeschke}
\affiliation{Research Institute for Interdisciplinary Science, Okayama University, Okayama 700-8530, Japan}

\author{Igor I. Mazin}
\email{imazin2@gmu.edu}
\affiliation{Department of Physics \& Astronomy, George Mason University, Fairfax, VA 22030, USA and
Quantum Science and Engineering Center, George Mason University, Fairfax, VA 22030, USA.}

\date{\today}

\begin{abstract}
Kagome lattice materials are an important platform for highly frustrated magnetism as well as for a plethora of phenomena resulting from flat bands, Dirac cones and van Hove singularities in their electronic structures. We study the little known metallic magnet CrRhAs, which belongs to a vast family of materials that include $3d$, $4f$ and $5f$ magnetic elements, as well as numerous nonmagnetic metals and insulators. Using noncollinear spin density functional calculations (mostly spin spirals), we extract a model magnetic Hamiltonian for CrRhAs. While it is dominated by an antiferromagnetic second nearest neighbor coupling in the kagome plane, the metallic nature of the compound leads to numerous nonzero longer range couplings and to important ring exchange terms. We analyze this Hamiltonian and find unusual ground states which are dominated by nearly isolated antiferromagnetic triangles that adopt 120$^\circ$ order either with positive or with negative vector chirality. 
We discuss the connection to the few known experimental facts about CrRhAs. 
Finally, we give a brief survey of other interesting magnetic members of this family of kagome compounds.
\end{abstract}

\maketitle

\section{Introduction}
Due to strong geometric frustration, antiferromagnetism on a kagome lattice is expected to yield novel properties such as classical or quantum spin liquids~\cite{10.1103/RevModPhys.89.025003,10.1126/science.aay0668,10.1088/0034-4885/80/1/016502,10.1038/nature08917}.
A good example with well-localized spin-1/2 copper magnetic moments forming a kagome lattice is herbertsmithite (\ce{ZnCu3(OH)6Cl2}) which is drawing a lot of interest since it was first synthesized in 2005~\cite{10.1021/ja053891p}. It has long been discussed as a quantum spin liquid candidate~\cite{10.1038/nature11659} but some kind of structural disorder plays a significant role~\cite{Barthelemy2022}. Other examples of kagome antiferromagnets which are proximate to or actually realize quantum spin liquids are kapellasite (\ce{ZnCu3(OH)6Cl2})~\cite{Fak2012,Iqbal2015}, Y-kapellasite (\ce{Y3Cu9(OH)19Cl8})~\cite{Barthelemy2019,Hering2022} and Zn-barlowite (\ce{ZnCu3(OH)6FBr})~\cite{Guterding2016,Feng2017}.

More recently, metallic kagome materials with or without magnetism have been studied intensively. 
A typical example is the intermetallic $T_mX_n$ kagome series ($T$ = Mn, Fe, Co; $X$ = Sn,  Ge; $m$:$n = 3$:$1, 3$:$2, 1$:$1$) with different kagome plane stackings~\cite{10.1038/s41563-019-0531-0}. Much effort has been put into studying their topological properties (Dirac Fermions and flat bands)~\cite{10.1038/s41563-019-0531-0,10.1038/s41467-021-25705-1,10.1038/nature25987}. Related kagome metals, such as \ce{YMn6Sn6}, exhibit a nontrivial topological Hall effect~\cite{10.1126/sciadv.abe2680,10.1038/s41467-021-23536-8,10.1103/PhysRevB.101.100405}, while another kagome metal, \ce{AV3Sb5}(A=K,Na,Cs) demonstrates a series of intriguing orders, including superconductivity~\cite{10.1103/physrevlett.127.177001,10.1038/s41586-022-04493-8,10.1103/physrevlett.126.247001,10.1038/s41586-021-04327-z}. 
Importantly, kagome planes in these systems are not magnetically frustrated, but they retain interesting electronic properties due to the special features of the kagome dispersion, Dirac point, van Hove singularities and flat band. Kagome metals with antiferromagnetic frustration have been little studied so far~\cite{Lacroix2010,Siddiqui2020}.

\begin{figure}[htbp]
    \centering
    \includegraphics[width=\columnwidth]{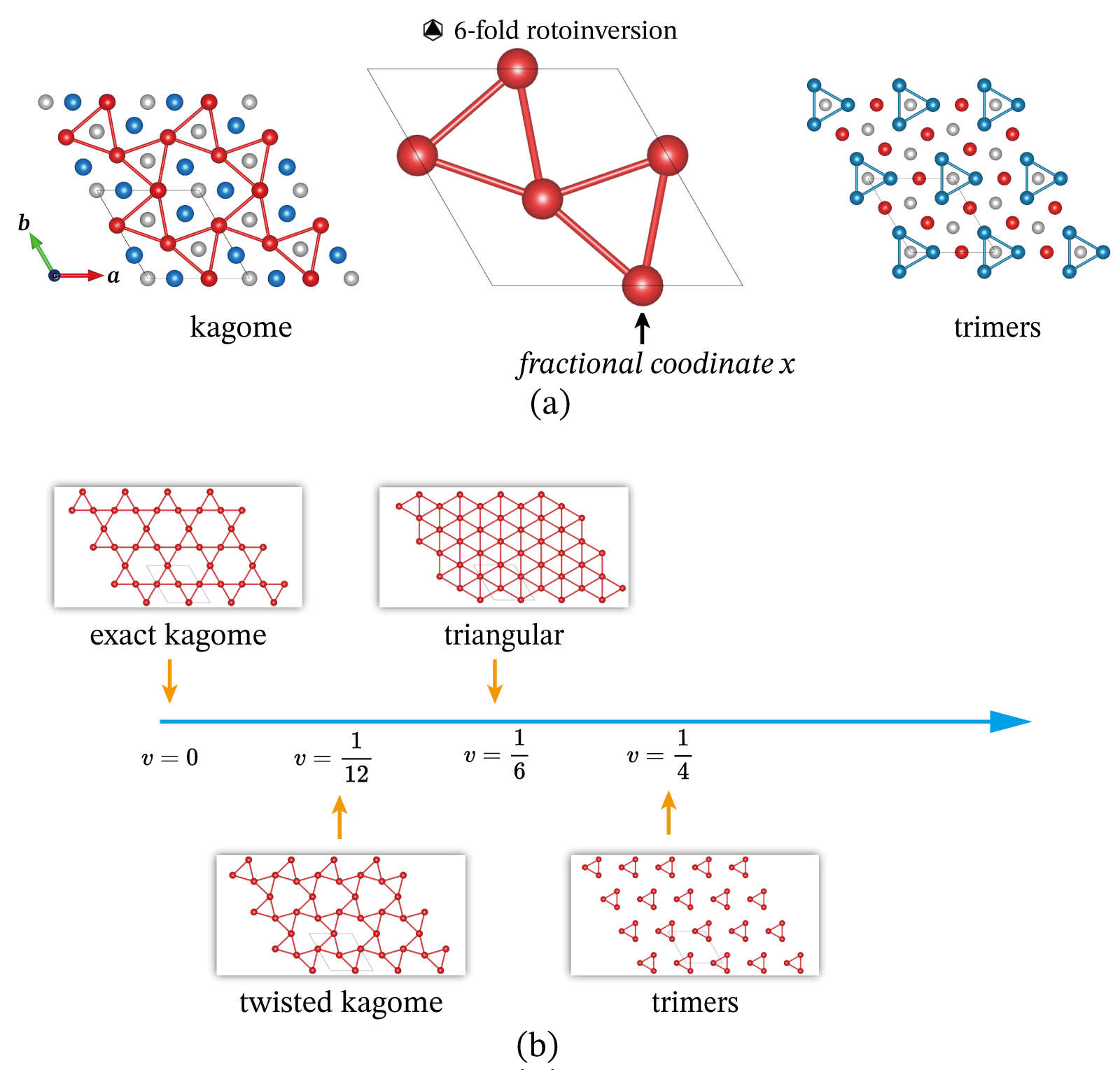}
    \caption{(a) Two 6-fold rotoinversion axes in the unit cell of the $P\bar{6}2m$ $XYZ$ compound (center). $x$ is the fractional coordinate of the $X$ atom indicated by an arrow. The left panel shows red atoms forming a kagome sublattice which is twisted by triangle rotations. The right panel shows blue atoms forming triangles. (b) The type of sublattice is uniquely determined by $v=x-1/2$. When $v=0$, the sublattice is exact kagome; When $|v|<1/6$, it is a twisted kagome; when $|v|>1/6$ it is trimers. The VESTA visualization program~\cite{Momma2011} was used to generate this figure.}
\label{fig:XYZ_kagome}
\end{figure}

Ideal kagome lattices are not uncommon, but relatively rare. An important point in this regard is that most unique properties of kagome magnets do not, actually, require an ideal kagome {\it geometry}, but rather an ideal kagome {\it connectivity}. In this respect, there is no difference at the nearest neighbor level between a perfect kagome lattice and the one twisted by triangle rotations, as shown in Fig.~\ref{fig:XYZ_kagome}. 
In this paper, we discuss a large family of compounds with the chemical formula $XYZ$ and space group $P\bar{6}2m$ (no. 189). Typically, they include two metal layers, $X$ and $Y$ with ligands $Z$ integrated into the two layers at a ratio 1:2. Thus, the structure can be understood as a stacking of $Y_3Z_2$ at $z=0$ and of $X_3Z$ at $z=0.5$.
Due to two 6-fold rotoinversion axes, the metal planes can be described as twisted kagome, where ligands sit in high-symmetry positions inside the metal planes. One metal sublattice, as discussed below, is only moderately deformed from the ideal kagome, while the other, incorporating two ligand atoms per unit cell, has metals fitted into trimers. The former subsystem is often magnetic, the latter usually not.

The elemental base for this crystallographic family, known by its prototype \ce{ZrNiAl}, where $X$=Zr, $Y$=Al, and $Z$=Ni, is large, and metallic layers can be formed by transition metals, lanthanides, actinides, alkaline earths, etc. 
Many of the compounds are magnetic, when $X$ forms the magnetic sublattice, and the resulting frustration is identical to the ideal kagome model. 
A known but little studied representative of this family is CrRhAs~\cite{Ohta1990,Kanomata1991,Kaneko1992,Ohta1995}. We have chosen it as an example to perform an extensive study of its in-plane and out-of-plane magnetic interactions, using density functional theory (DFT) calculations. We use the energies of spin spirals to extract the important parameters of a Heisenberg plus ring exchange Hamiltonian. We find that the second inplane exchange interaction clearly dominates over the first, leading to a spiral magnetic ground state.

\section{{$XYZ$} compounds with \texorpdfstring{$P\bar{6}2m$}\bf{} structure}

In $XYZ$ with $P\bar{6}2m$ space group, both $X$ and $Y$ sublattices are characterized by one distortion parameter $v=x-\frac{1}{2}$, where $x$ is the coordinate of the $3g$ ($3f$) Wyckoff position. The $3g$ and $3f$ positions differ only in the $z$ coordinate, 1/2 and 0, respectively. The $Z$ ions occupy two sublattices, $1b$ in the $3g$ plane, and $2c$ in the $3f$ plane. 
Increasing the absolute value of the distortion parameter $|v|$ makes the equilateral $X$ triangles grow and rotate in the $X_3Z$ plane.
This takes the $X$ sublattice from an ideal kagome lattice at $v=0$ via a kagome lattice with rotated triangles and deformed hexagons for $0<|v|<\frac{1}{6}$ and a perfect triangular lattice at $|v|=\frac{1}{6}$ to trimers for $|v|>\frac{1}{6}$. 
At increasing $|v|$, the triangular lattice of $Z$ at the $1b$ position is enclosed by ever smaller $X$ triangles.
The $v$ parameter has the same effect in the $Y_3Z_2$ plane with the difference that here, the $Z$ in the $2c$ position form a honeycomb lattice.
Thus, the connectivity in the two metal sublattices, $X$ and $Y$, is different, which dramatically affects their magnetic properties. The compact triangles
in the $Y_3Z_2$ plane tend to have considerable covalent bonding, and no, or little magnetism.
The $X$ ions, in contrast, form only a moderately twisted kagome lattice (minimizing the Coulomb interaction with the ligand in the center), and are likely to have magnetism which can be frustrated in case of antiferromagnetic interactions.

We have inspected the $P\bar{6}2m$ $XYZ$ compounds on the materials project website~\cite{10.1063/1.4812323} and organized some potentially magnetic ones into convenient tables, shown in Ref.~\cite{SM}, Tables~S1 to S3. 
We found a number of $XYZ$ compounds with significant magnetism, where magnetic kagome atoms can be Ce, Cr, Eu, Fe, Gd, Mn, Np, Pu, or U. 

\section{C\lowercase{r}R\lowercase{h}A\lowercase{s}}
We used a projector augmented wave basis as implemented in the Vienna ab initio simulation package (VASP)~\cite{10.1103/PhysRevB.54.11169,10.1016/0927-0256(96)00008-0,10.1103/PhysRevB.47.558} to perform non-collinear magnetic calculations -- when needed, with individual constraints. We use all electron calculations with the full potential local orbital (FPLO) basis~\cite{Koepernik1999} to plot band structure and Fermi surface. The generalized gradient approximation(GGA) in the Perdew-Burke-Ernzerhof variant(PBE)~\cite{10.1103/PhysRevLett.77.3865} 
was used as the exchange-correlation potential.

We base our calculations on the crystal structure of CrRhAs determined by Deyris {\it et al.}~\cite{Deyris1979} (ICSD 43919)
with $a=b= 6.384(1)$\,{\AA} and $c=3.718(1)$\,{\AA}.
The internal atomic positions of CrRhAs were relaxed in VASP while keeping lattice parameters fixed. The optimized structural parameters are shown in Table \ref{tab:optimization}, and will be used from now on.
\begin{table}[htbp]
    \centering
    \caption{GGA optimized fractional coordinates of CrRhAs.\label{tab:optimization}}
    \begin{tabularx}{\columnwidth}{*{5}{X}}
    \toprule
        Atom & Wyckoff & x & y & z \\ \midrule
        Cr & $3g$ & 0.6017 & 0 & 0.5  \\ 
        Rh & $3f$ & 0.2641 & 0 & 0  \\
        As1 & $1b$ & 0 & 0 & 0.5  \\ 
        As2 & $2c$ & 1/3 & 2/3 & 0  \\
    \bottomrule
    \end{tabularx}
\end{table}

\section{Magnetic pattern analysis}

\begin{figure}[htbp]
	\centering
	\includegraphics[width=0.9\columnwidth]{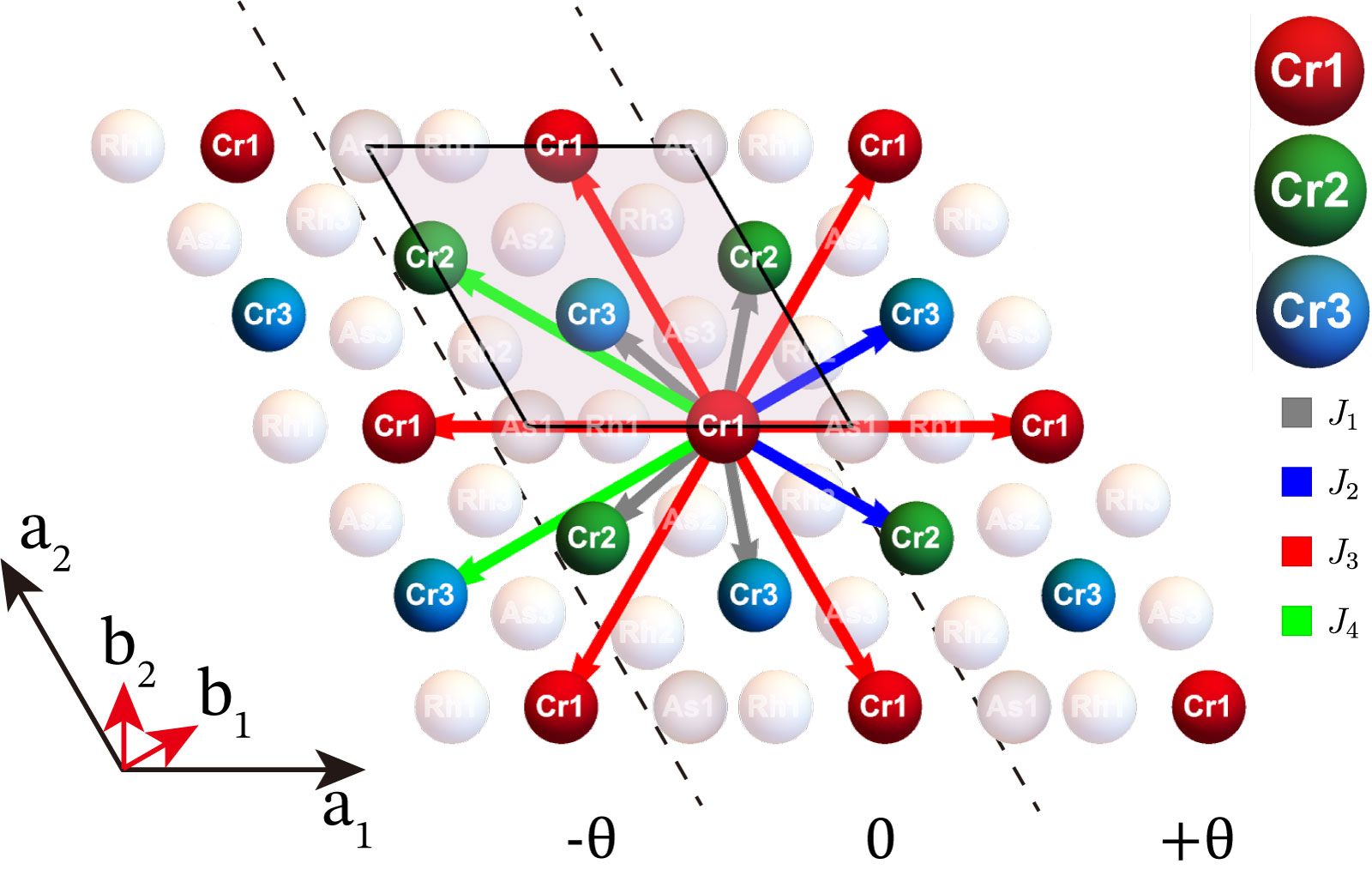}
	\caption{1st to 4th in-plane nearest neighbors in the Cr kagome sublattice in CrRhAs, indicated by arrows of different colors. The dashed lines cut the lattice into stripes corresponding to different spiral angles. Red, blue and green atoms are Cr1, Cr2, Cr3 respectively.}
	\label{fig:q00strip}
\end{figure}

\begin{figure}[htbp]
	\centering
	\includegraphics[width=\columnwidth]{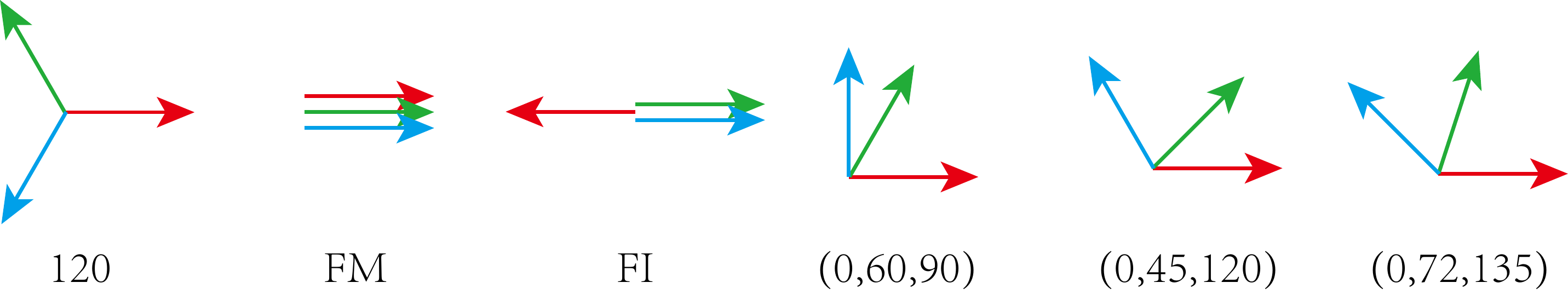}
	\caption{
		6 spin configurations within a unit cell that considered in spin spiral calculations. Red, Green, Blue arrows indicates moment directions of Cr1,Cr2,Cr3
	}
	\label{fig:q00SpinConfiguration}
\end{figure}

\begin{table}[htbp]
	\centering
	\caption{$\phi(1), \phi(2), \phi(3)$ (in degrees) for the six different spin spirals\label{tab:angles}}
	\begin{tabularx}{\columnwidth}{p{0.4\columnwidth}*{3}{p{0.2\columnwidth}}}
		\toprule
		spin configuration   & $\phi(1)$   & $\phi(2)$   & $\phi(3)$    \\
		\midrule
		120    & 0   & 120 & 240   \\
		FM     & 0   & 0   & 0     \\
		FI   & 180 & 0   & 0     \\
		(0,60,90)  & 0   & 60  & 90    \\
		(0,45,120) & 0   & 45  & 120  \\
		(0,72,135) & 0   & 72  & 135  \\
		\bottomrule
	\end{tabularx}
\end{table}

\begin{figure}[htbp]
	\centering
	\includegraphics[width=0.35\columnwidth]{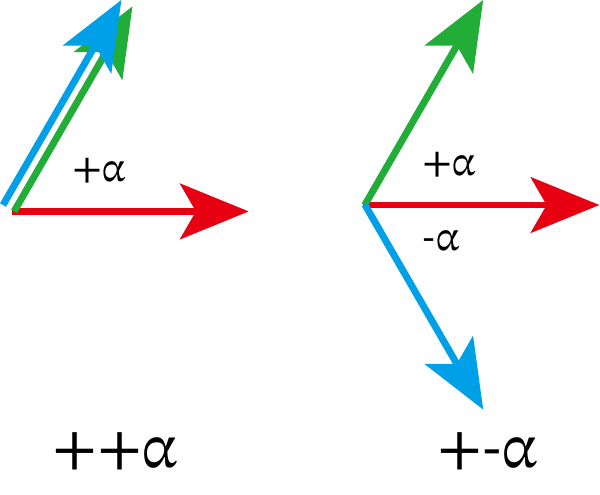}
	\caption{Spin configuration of noncollinear ++$\alpha$ and +-$\alpha$. Red, green and blue vector are spin moments of Cr1, Cr2 and Cr3 sublattice respectively.}
	\label{fig:+++-alpha}
\end{figure}

In the following, we will infer magnetic couplings between Cr atoms based on energies of different magnetic patterns on the Cr sublattices. 
In Fig.~\ref{fig:q00strip}, we define the three sublattices Cr1, Cr2, Cr3 of the kagome lattice.
Arrows of different colors indicate the 1st to 4th in-plane nearest neighbor Cr atoms to a reference Cr1 atom. Each arrow associated with a Heisenberg term ${J_n}{{\bf{s}}_i} \cdot {{\bf{s}}_j}$ where $J_n$ means nth in-plane nearest Heisenberg coupling and $\textbf{s}_{i=1,2,3}$ means normalized spin operators for Cr1, Cr2, Cr3. The polar angles of moments are assumed to be $\phi(1),\phi(2),\phi(3)$ for Cr1, Cr2, and Cr3, respectively.

We will discuss spin spirals propagating along different directions and with 6 cases of spin configurations within a unit cell defined as 120, FM, FI, (0,60,90),(0,45,120), (0,72,135) as shown in Fig.~\ref{fig:q00SpinConfiguration}. 
The $\phi(1),\phi(2),\phi(3)$ for each case are listed in Table.~\ref{tab:angles}.
Besides, we also discuss periodic magnetic patterns with varying angle $\alpha$ as defined in Fig.~\ref{fig:+++-alpha}

The total energy of a magnetic pattern on Cr sublattices can be written as 
\begin{equation}
	H = E_0 + H_{\mathrm{Heisenberg}} + H_{\mathrm{ring}}
\end{equation}
where $H_{\mathrm{Heisenberg}}$ is defined as
\begin{equation}
H_{\mathrm{Heisenberg}} = \sum_{i<j} J_{ij} \mathbf{s}_i \cdot \mathbf{s}_j \label{eq2}
\end{equation}
Here, the $\mathbf{s}_i = \frac{1}{S}\mathbf{S}_i$ are  magnetic moment vectors normalized to 1. We write the exchanges in the kagome plane as $J_{ij}=J_1, J_2, \dots$ where $1, 2, \dots$ correspond to increasing bond length (i.e. coordination shell). Exchange between the kagome layer and the next layer above and below are $J_{ij}=J_0', J_1', J_2', \dots$, again sorted by distance ($J_0'$ is straight up or down).  $J_{ij}=J_0'', J_1'', J_2'', \dots$ are bonds connecting second layers, and so on.

$H_{\text{ring}}$ is the ring exchange on Cr triangle defined as
\begin{equation}
{H_{{\text{ring}}}} 
= { {{L_1}\sum\limits_{\vartriangle_1}  {{{\bf{s}}_i} \cdot {{\bf{s}}_j}} }+{{L_2}\sum\limits_{\vartriangle_2}  {{{\bf{s}}_i} \cdot {{\bf{s}}_j}} } }
\end{equation}
where
\begin{equation}
    \sum\limits_\mathrm{\vartriangle_1 or \vartriangle_2}=({{{\bf{s}}_1} \cdot {{\bf{s}}_2}} )( {{{\bf{s}}_2} \cdot {{\bf{s}}_3}} ) +
    ( {{{\bf{s}}_1} \cdot {{\bf{s}}_3}} )( {{{\bf{s}}_3} \cdot {{\bf{s}}_2}} )+ ( {{{\bf{s}}_2} \cdot {{\bf{s}}_1}} )( {{{\bf{s}}_1} \cdot {{\bf{s}}_3}})
\end{equation}
Here, sum over $\vartriangle_1$ or $\vartriangle_2$ means $\mathbf{s}_{i=1,2,3}$ are three moments of Cr1, Cr2, Cr3 that form 1st or 2nd nearest Cr triangle. $L_1$ and $L_2$ are corresponding triangle ring exchange strengths.

With the above definitions, we can rewrite the total energy in Eq.\ref{eq:total_energy} as
\begin{equation}
	\label{eq:total_energy}
	E = {E_0} +  L_1  D_1 +  L_2  D_2 + \left(\sum\limits_i {{J_i} {C_i}}   + \sum\limits_i {{J_i}' {{C_i}'}}  +  \cdots  \right)
\end{equation}
where $C_i$, $C_i'$, $D_1$, $D_2$ are summations of terms like $\mathbf{s}_i\cdot\mathbf{s}_j$ and $(\mathbf{s}_i\cdot\mathbf{s}_j)(\mathbf{s}_j\cdot\mathbf{s}_k)$ that depend on magnetic patterns.

\begin{figure}[htbp]
	\centering
	\includegraphics[width=\columnwidth]{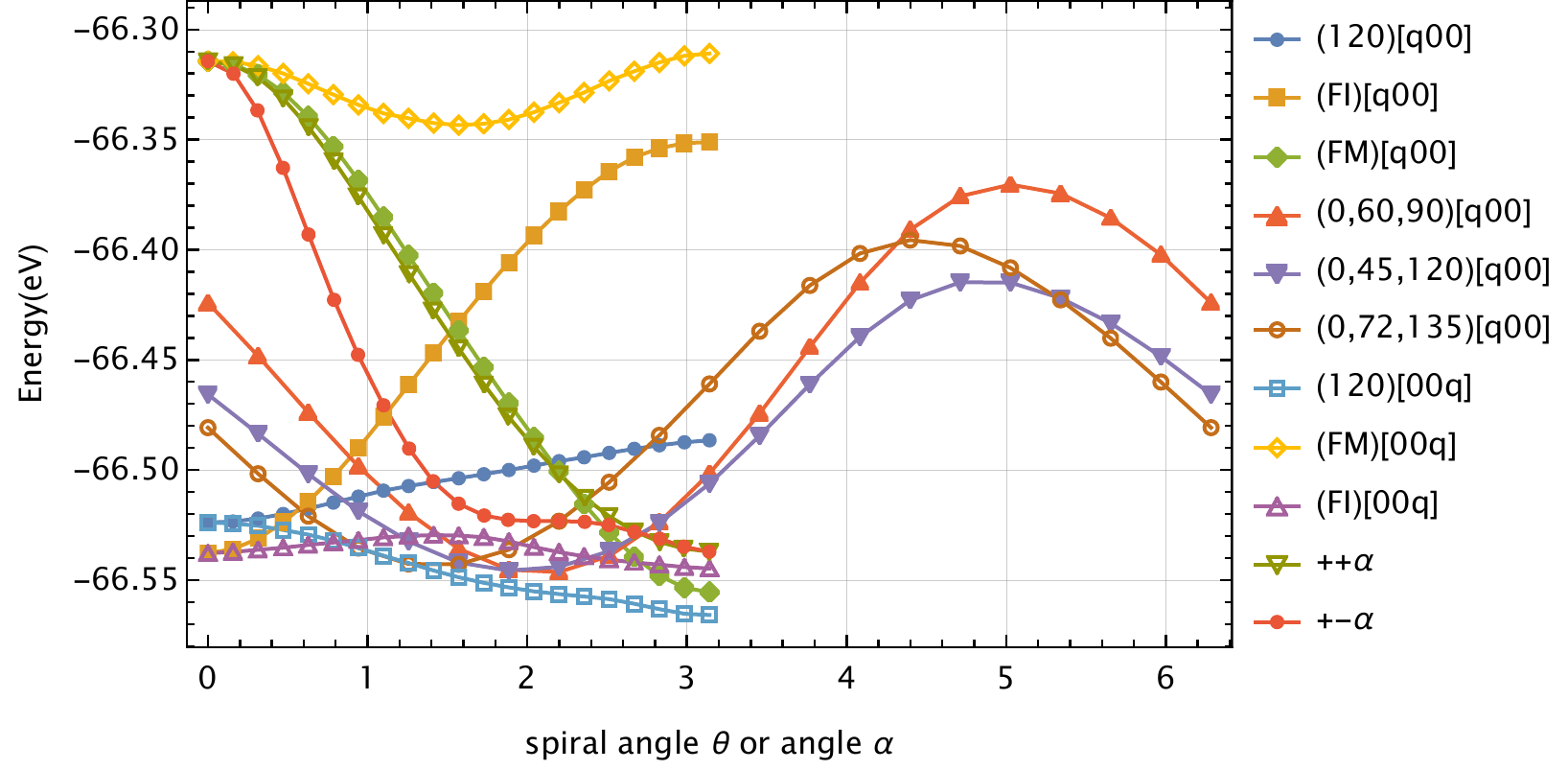}
	\caption{Energy curves of 6 $(q,0,0)$ spirals, 3 $(0,0,q)$ spirals, ++$\alpha$ and +-$\alpha$.}
	\label{fig:q00_00q_alpha_curve}
\end{figure}

\subsection{Spin spiral configurations}
Spin spirals were modeled using the generalized Bloch theorem~\cite{10.1088/0953-8984/3/44/004}, as implemented in VASP. We calculated various spin spiral configurations for CrRhAs in order to get consistent information about the magnetic coupling between Cr atoms in CrRhAs.
A spin spiral is determined by a propagation vector $\textbf{q}$ within the first Brillouin zone of the reciprocal space lattice. 
We mainly considered $(q_x,0,0)$ and $(0,0,q_z)$ spirals.

The $(q_x,0,0)$ spirals are propagating along reciprocal $\mathbf{b}_1$ direction which is shown in Fig.~\ref{fig:q00strip}. 
To understand the connectivity in these spirals, we can divide the lattice into stripes (dashed lines in Fig.~\ref{fig:q00strip}) running along the unit cell vector $\mathbf{a}_2$ which is perpendicular to $\mathbf{b}_1$.  
From one stripe to the next, all moments are rotated by the angle $\theta_x=\mathbf{q}\cdot \mathbf{a}_1$.
There is an additional freedom of choosing Cr moment directions $\phi(1),\phi(2),\phi(3)$ within the unit cell, which generates different spin configurations.
We considered six types of $(q_x,0,0)$ spin spirals, labeled as 
$(120)[q00]$, $({\mathrm{FM}})[q00]$, $({\mathrm{FI}})[q00]$, $(0,60,90)[q00]$, $(0,45,120)[q00]$, $(0,72,135)[q00]$ and corresponding $(0,0,q_z)$ spirals labeled by replacing $[q00]$ with $[00q]$.
For $(0,0,q_z)$, a spiral propagates along the $\mathbf{a}_3$ direction, which is much simpler. All moments in a horizontal plane have the same spiral angle, and in the next plane along the $\mathbf{a}_3$ direction, they rotate by $\theta_z=\mathbf{q}\cdot \mathbf{a}_3$. 
Explaining the energetics of such spirals requires out of plane exchange couplings. 

Simple, if tedious, calculation renders a Heisenberg Hamiltonian which is a linear form in $J$s and $L$, with the coefficients $C_i$ and $D$ for general $(q_x,0,q_z)$ spin spiral, as shown in Ref.~\cite{SM} (Table.~S5). 
For each spiral of the 120, FI, and FM cases, we performed spiral total energy calculations from spiral angle 0 to $\pi$, because their energy is symmetric with respect to spiral angle $\pi$ up to $J_4$. 
For the other cases, $(0,60,90)$, $(0,45,120)$, $(0,72,135)$, we calculated the full spiral angle range 0 to $2\pi$.

\subsection{Noncollinear periodic calculations}

We also consider simple periodic cases where only the three Cr sublattices have different noncollinear spin directions, as shown in Fig.~\ref{fig:+++-alpha}, and where we vary the angle $\alpha$. 
Taking the spin direction of Cr1 as reference, in the ++$\alpha$ case the spin directions of Cr2 and Cr3 are rotated by the same angle $+\alpha$, while in the +-$\alpha$ case, the spin direction of Cr2 and Cr3 are rotated by $+\alpha$ and $-\alpha$, respectively. 
The energy versus $\alpha$ curves and their fittings to different orders of cosine are shown in Ref.~\cite{SM} (Fig.~S2).

If the magnetic interaction is dominated by Heisenberg type contributions, we would expect that a simple $\cos(\alpha)$ will fit the ++$\alpha$ curve well. 
As we can see from Fig.~S2 (see Ref.~\cite{SM}),
the fit is indeed not bad with single $\cos(\alpha)$, but after adding $\cos(2\alpha)$ it becomes much better.
For the +-$\alpha$ case, we would expect $\cos(\alpha)$ plus $\cos(2\alpha)$ will fit well. However, it turns out that fitting to the order of $\cos(2\alpha)$ still has a large discrepancy with the calculated energies, and by adding a $\cos(3\alpha)$ term, two curves immediately snap together. So it is clear that explaining the +-$\alpha$ curve needs the $\cos(3\alpha)$ term.

One probable explanation is that there is ring exchange between three Cr atoms. For the spins $\mathbf{s}_{1}$,  $\mathbf{s}_{2}$, $\mathbf{s}_{3}$ of $\text{Cr}_{i=1,2,3}$, a ring exchange interaction is proportional to $\left( {{{\bf{s}}_1} \cdot {{\bf{s}}_2}} \right)\left( {{{\bf{s}}_2} \cdot {{\bf{s}}_3}} \right) + \left( {{{\bf{s}}_1} \cdot {{\bf{s}}_3}} \right)\left( {{{\bf{s}}_3} \cdot {{\bf{s}}_2}} \right) + \left( {{{\bf{s}}_2} \cdot {{\bf{s}}_1}} \right)\left( {{{\bf{s}}_1} \cdot {{\bf{s}}_3}} \right)$. In the +-$\alpha$ case, this will introduce $\cos(\alpha)\cos(2\alpha)$ and $\cos(\alpha)^2$ terms, and $\cos(\alpha)\cos(2\alpha)$ is equivalent to $\frac{1}{2}\big(\cos(3\alpha) + \cos(\alpha)\big)$, so that $\cos(3\alpha)$ emerges as soon as we consider ring exchange.
For the ++$\alpha$ case, the ring exchange term introduces additional $\cos(\alpha)^2$ contributions which also improves the fit.

\begin{figure*}[htbp]
	\centering
	\includegraphics[width=\textwidth]{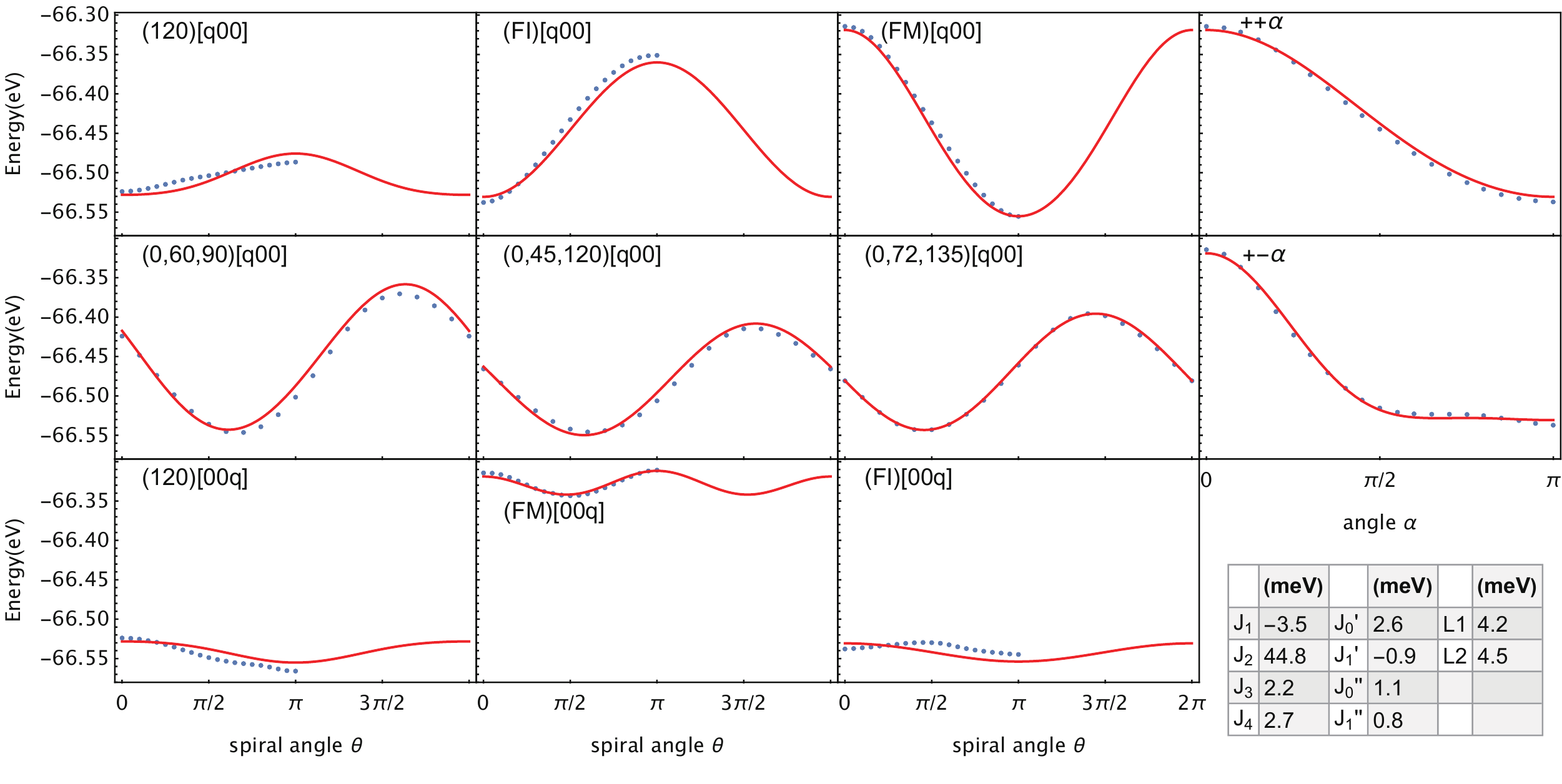}
	\caption{Least square fit of six $(q_x,0,0)$ spirals,  three $(0,0,q_z)$ spirals, ++$\alpha$ and +-$\alpha$ to the Heisenberg Hamiltonian including four in-plane neighbors, two  neighbors in the first and second interplanar interactions, and the nearest neighbor ring exchange, as discussed in the text. Blue symbols are calculated DFT energies. Red curves are the least squares fit with the parameters shown on the right.}
	\label{fig:q00_00q_alpha_fitting}
\end{figure*}

\subsection{Fitting all energy curves}
Energies of six $(q_x,0,0)$ spirals, three $(0,0,q_z)$ spirals, ++$\alpha$ and +-$\alpha$ are shown in Fig.~\ref{fig:q00_00q_alpha_curve}. 
Note that we have intentionally shifted the curves of ++$\alpha$ and +-$\alpha$ up to align with $({\mathrm{FM}})[q00]$ and $({\mathrm{FM}})[00q]$ at $\theta=0$, because due to the internal realization of VASP, energies of noncollinear and spiral calculations have a constant shift.
From Fig.~\ref{fig:q00_00q_alpha_curve} The $120[00q]$ with $\theta=\pi$ has the lowest energy. This means that the interlayer coupling is AFM, and moments on Cr atoms on the same triangle tend to form the 120 degrees with each other.

Now we have 9 spiral curves plus 2 noncollinear curves with variable angle as shown in Fig.~\ref{fig:q00_00q_alpha_curve}, which can be fit to Eq.~\ref{eq:total_energy}, using Table~S5 (see Ref.~\cite{SM}). 
It appears that only a subset of $J$s are linearly independent; furthermore, some longer-range couplings, while they can formally be extracted by the fit, come out very small and improve the fit only marginally. A choice of $E_0,J_1,J_2,J_3,J_4,J_0',J_1',J_0'',J_1'',L$ as a physically meaningful Hamiltonian gives good overall fits as shown in Fig.~\ref{fig:q00_00q_alpha_fitting}.

From this fitting result, we found that $J_2$ is the dominant exchange interaction. It is an antiferromagnetic coupling, and interestingly it is 10 times larger than the the second largest ferromagnetic exchange interaction $J_1$. What is more, $J_3, J_4$ are also of the same order of magnitude as $J_1$. The nearest and next nearest interlayer coupling is antiferromagnetic. Thus, we find a Heisenberg Hamiltonian with clearly dominating antiferromagnetic interactions, in agreement with the fact that experimentally, CrRhAs was found to order antiferromagnetically with a Neel temperature of $T_{\rm N}=165$\,K~\cite{Ohta1990}. Considering the hierarchy of exchange couplings, we expect the $J_2$ triangles to order in a 120 degree state. The second largest ferromagnetic $J_1$ couplings cannot be exactly satisfied because of $J_1$-$J_2$-$J_1$ triangles, and they already introduce some frustration. The smaller in-plane couplings $J_3$ and $J_4$ also contribute to frustration. Interestingly, even though the interlayer distances of CrRhAs are small, interlayer exchange is much smaller than in-plane exchange, and the material is magnetically rather two-dimensional.  

Furthermore, the ring exchange term is indispensable for good fits of the DFT energies and is substantial at 12\% of the dominant exchange interaction. Note that we can directly compare Heisenberg and ring exchange terms as we are using unit moments. As shown in Ref.~\cite{SM}, Fig.~S3, 
without ring exchange, there are discrepancies between fitted and original data curves as large as \SI{20}{meV} for $(120)[q00]$ and $(120)[00q]$ at $\theta=0$, and similarly for +-$\alpha$.

\section{Discussion of the emerging Hamiltonian}
\begin{figure}[htbp]
	\includegraphics[width=\columnwidth]{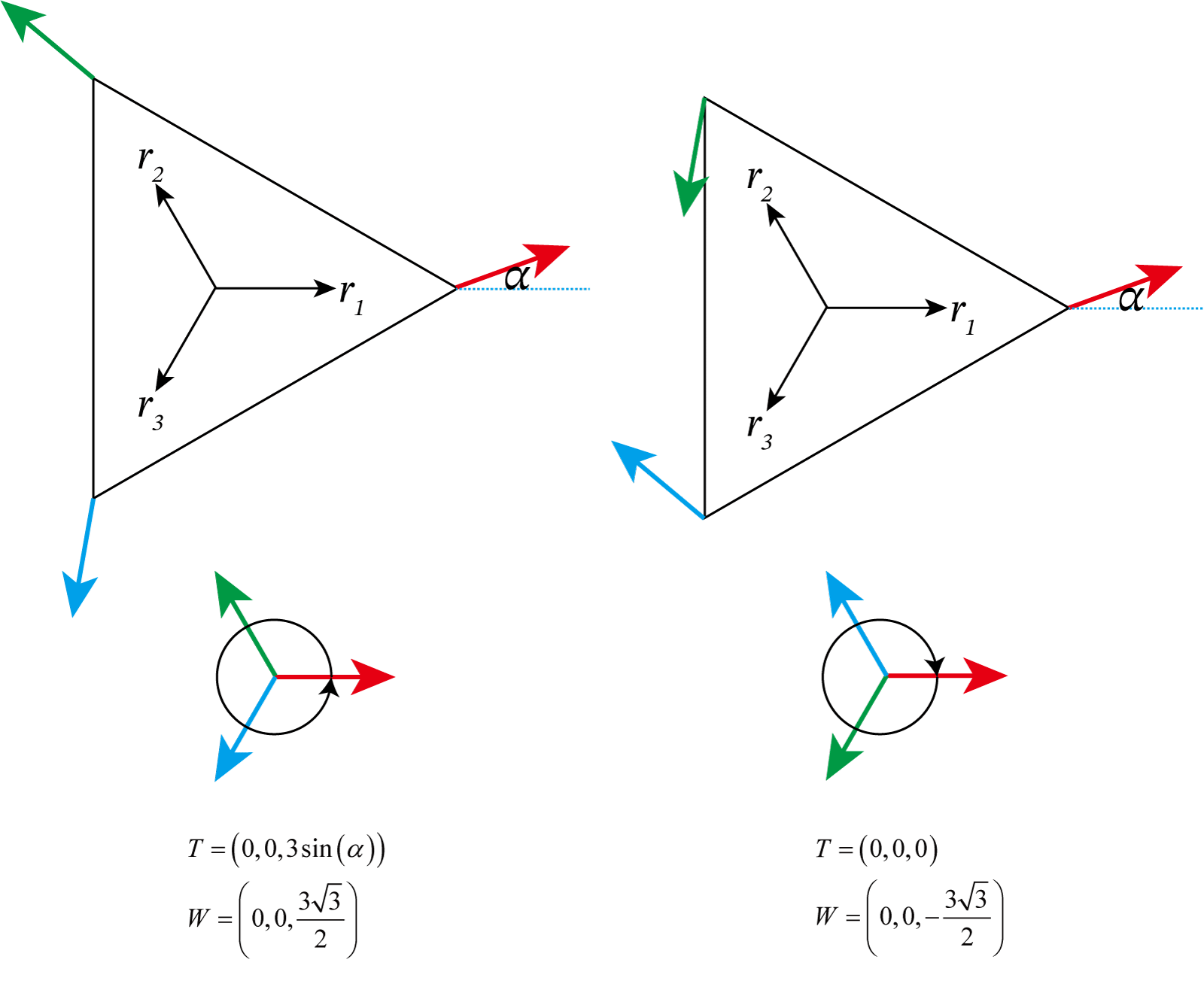}
	\caption{Two topologically different 120$^\circ$ magnetic patterns on a triangle. Red, green and blue vector are spin moments of Cr1, Cr2 and Cr3 respectively. Note that the left pattern has an $\alpha$ dependent, usually nonzero toroidal moment $T=3\sin{\alpha}$ (assuming the black vectors $r_i$ and magnetic moments have unit lengths), which can be continuously varied from $3$ to $-3$ by rotating the spin space with respect to the coordinate space, while the right pattern has always $T=0$, and that is not affected by the spin-space rotation. The left pattern has positive, the right negative vector chirality. 
	}
	\label{T}
\end{figure}

\begin{figure}[htbp]
    \includegraphics[width=.97\columnwidth]{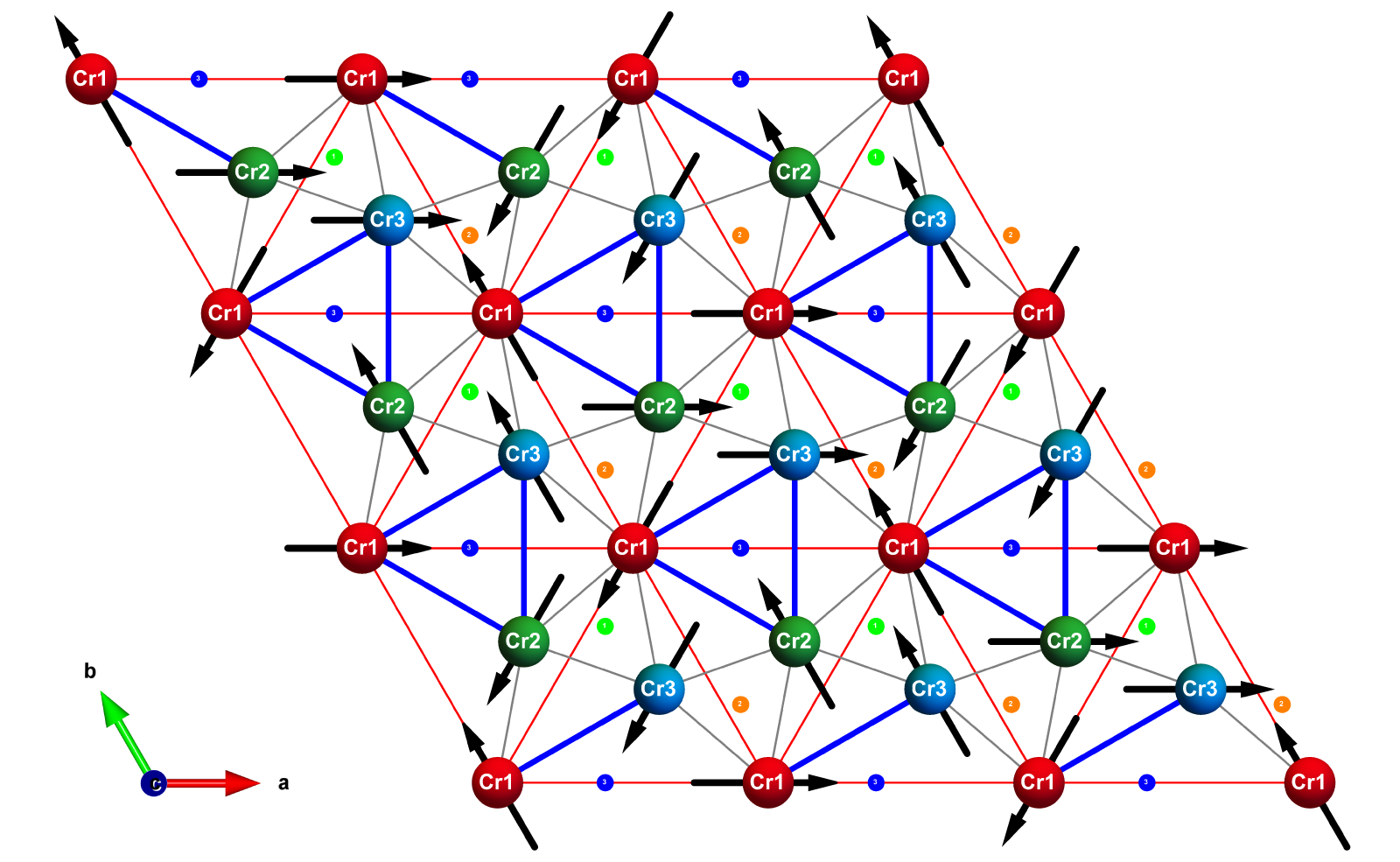}\\
    \includegraphics[width=.97\columnwidth]{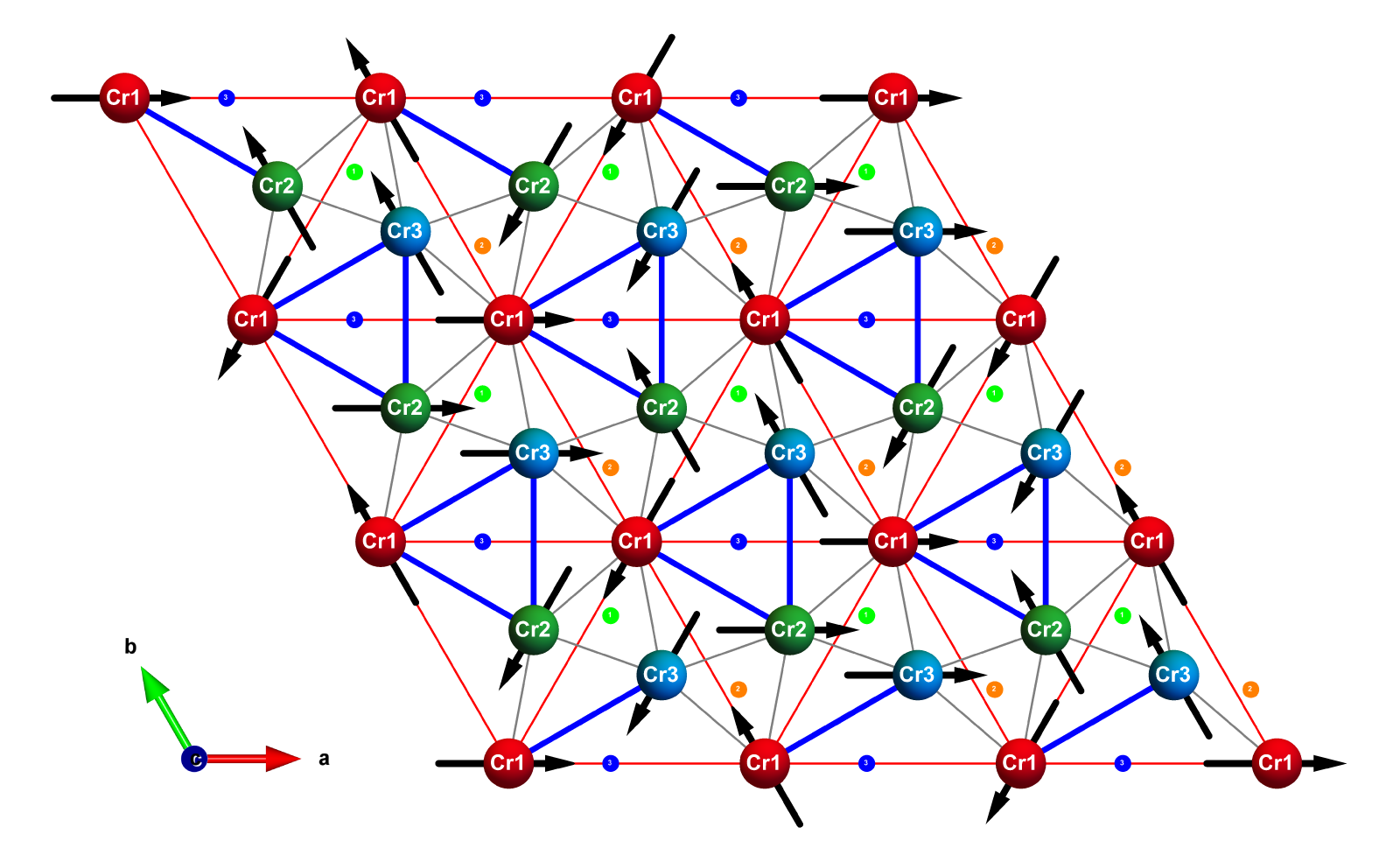}
    \caption{Two possible 2D ground state of the reduced magnetic Hamiltonian including $J_{1}$ to $J_4$ interactions (note that the ground state is highly degenerate). 
    The blue bonds indicate the strongest exchange coupling (AF) in the system ($J_2$), which generates 120$^\circ$ trimers, the grey bonds the nearest neighbor triangles, and the red one effective inter-trimer AF interaction, numerically equivalent to $J_3-J_1$. Note that while in the ground state the blue trimers are ordered in a 120$^\circ$ fashion, and the red bonds network also assumes a 120$^\circ$ order, the two orders, in this model, are not correlated and may have different ordering planes and toroidal moments. The blue balls indicates centers of the blue trimers, and the green and orange ones centers of the n.n. triangles. The top diagram corresponds to an order with negative vector chirality on the blue triangles, the bottom one to one with positive vector chirality. 
    }
\label{vesta}
\end{figure}

The Hamiltonian derived in the previous section is quite unusual. First, it is dominated by the large AF 2nd nearest-neighbor interaction (blue bonds). These bonds form isolated triangles, all oriented in the same way. Each triangle, obviously, orders in a 120$^\circ$ fashion, and is formed by the three different Cr, Cr1, Cr2 and Cr3. Let us first for simplicity assume an XY model, so all spins lie in the $ab$ plane (Fig. \ref{vesta}).
There are two different ways to produce this order, illustrated in Fig. \ref{T}, differing by the sign of their vector chirality $\mathbf{W}=\mathbf{M}_1\times\mathbf{M}_2+\mathbf{M}_2\times\mathbf{M}_3+\mathbf{M}_3\times\mathbf{M}_1$ on the dominant $J_2$ triangles. We use positive or negative vector chirality to distinguish between the two states (see Fig.\ref{T}). Toroidal moment $\mathbf{T}=\sum_{i=1}^3\mathbf{r}_i\times\mathbf{M}_i$ is usually nonzero for the state with positive vector chirality while it is always zero for the state with negative vector chirality.
After one or the other types is selected, each blue triangle is fully determined by one of its spins (let's say, by the Cr1 spin. Then the lattice of the blue triangles is equivalent to a triangular lattice shown in red. 

Let us now determine the effective Hamiltonian for this lattice: consider two blue triangles shifted along $a$. The Cr1 on the left is connected to Cr2$'$ and Cr3$'$ on the right, where ``$'$'' means the  atoms from the right triangle. The corresponding contribution to energy is $J_1\mathbf{S}_1\cdot (\mathbf{S}_2'+\mathbf{S}_3')=-J_1\mathbf{S}_1\cdot \mathbf{S}_1'$. This, the FM n.n. interaction gives rise to an AF interaction for this green bond, which needs to be added to the, also AF, $J_3$. 

Let us now evaluate the interaction along $b$. By the same token, the coupling between the corresponding Cr2$'$ atoms is also $J_{\rm eff}=J_3-J_1$. Since Cr2's are just the Cr1's in the same black triangle rotated by $2\pi/3$, it is the same as adding $J_{\rm eff}$ along $b$ for the Cr1 atoms (note that in principle we could rotate spins in the opposite directions when shifting along $b$ compared to shifting along $a$, but that would have been energetically unfavorable). 

Thus, we get a unique ground state, where the spins on the blue sublattice, which are of three different colors, form a 120$^\circ$-lattice, and each color within itself also forms 120$^\circ$-lattices. 
Next, let us look at the black triangles. Their centers are denoted by green and orange balls. They form a perfect honeycomb lattice, but it partitions (as the honeycomb lattice is bipartite)  into two (triangular again) subsets (green and orange), one sporting FM triangles and the other 120$^\circ$-triangles. Note that in terms of the {\it toroidal} moments it is reversed: the former subset has zero toroidal moments, while the latter non-zero ones. Note that the $J_4$ interaction, comparable with $J_3$ (but considerably smaller than $J_{\rm eff}=J_3-J_1$) is also satisfied as well as it is possible for a triangular lattice, $i.e.$, with 120$^\circ$ angles.

If we now select the other pattern with a positive vector chirality on the dominant $J_2$ (blue) triangles, we end up with and alternative structure, strictly degenerate with the first one (Fig. \ref{vesta}, bottom).

\begin{table*}[htbp]
    \centering
    \caption{Toroidicity and chirality in the spin configuration with negative vector chirality on the dominant $J_2$ triangles, as discussed in the main text. For the purpose of this table, the two ordering planes are assumed parallel, $\mathbf{\Omega}\parallel\boldsymbol{\omega}$, so that the scalar chirality is always zero.  \label{tab:achiral}}
    \begin{tabular}{l|l|l|c|c|c|c}
    \toprule
label&triangles                         &connectivity& effective coupling&net moment& toroidicity&vector chirality  ($\parallel z$)\\ \midrule
a    &nearest neighbors                 &honeycomb   &$J_1$               & &&\\
     &{\ \ \ \ sublattice 1}&&                                  & 3       & 0          &0\\
     &{\ \ \ \ sublattice 2}&& 
     & 0       & [-3,3], $\langle ...\rangle$=0& $3\sqrt{3}/2$\\\hline
b    &2nd neighbors                     &trimers     &$J_2$               & 0       & 0          & $-3\sqrt{3}/2$  \\\hline 
c    &centers of the ``b'' triangles$^1$&triangular  & $J_3-J_1$          & 0       & [-3,3], $\langle ...\rangle$=0& $\langle ...\rangle$=0  \\\hline
d    &4th neighbors                     &trimers     &$J_4$               & 0       & 0          & $-3\sqrt{3}/2$  \\
    \bottomrule
    \end{tabular}\\
\end{table*}

\begin{table*}[htbp]
    \centering
    \caption{Same as Table~\protect\ref{tab:achiral}, for the spin configuration with positive vector chirality on the dominant $J_2$ triangle. \label{tab:chiral}}
    \begin{tabular}{l|l|l|c|c|c|c}
    \toprule
label&triangles                         &connectivity& effective coupling&net moment& toroidicity&vector chirality  ($\parallel z$)\\ \midrule
a    &nearest neighbors                 &honeycomb   &$J_1$               &&&\\
     &{\ \ \ \ sublattice 1}&&
     & 3       & 0          & 0\\
     &{\ \ \ \ sublattice 2}&&
     & 0       & 0          & $-3\sqrt{3}/2$\\\hline
b    &2nd neighbors                     &trimers     &$J_2$               & 0       & [-3,3], $\langle ...\rangle$=0& $3\sqrt{3}/2$  \\\hline 
c    &centers of the ``b'' triangles$^1$&triangular  & $J_3-J_1$          & 0       & [-3,3], $\langle ...\rangle$=0& $\langle ...\rangle$=0  \\\hline
d    &4th neighbors                     &trimers     &$J_4$               & 0       & [-3,3], $\langle ...\rangle$=0& $3\sqrt{3}/2$  \\
    \bottomrule
    \end{tabular}\\
    $^1$ equivalent to the 3rd neighbor triangles
\end{table*}

To summarize, the effective model can be mapped onto a triangular lattice where each site is characterized by a Heisenberg-spin variable (let us call it $\mathbf{\Sigma}$, which shows the spin direction of the selected corner (Cr1, in this case), and another unit-length axial vector,  $\mathbf{\Omega}$, showing the sense of the rotation in a given blue triangle (i.e., rotations from Cr1 to Cr2 to Cr 3 by $2\pi/3$ proceed around the axis $\mathbf{\Omega}$, and the sense of the rotation is given by the sign of  $\mathbf{\Omega}$). The interaction between the effective spins $\mathbf{\Sigma}$ is also Heisenberg, AF, and much smaller that the interactions inside a trimer, and leads to a standard triangular Heisenberg order, which can also be characterized by a Heisenberg spin variable, $\mathbf{S_0}$, which can be selected as the value $\mathbf{\Sigma_0}$ at the origin, and another unit-length rotation vector, $\boldsymbol\omega$. At the end, the entire long-range magnetic order can be described by one Heisenberg spin $\mathbf{\Sigma_0}$ and two axial rotational vectors $\mathbf{\Omega}$ and $\boldsymbol{\omega}$. This is to be contrasted with the less-degenerate standard triangular lattice, which can be uniquely described by the origin spin and one rotational vector. 

Consideration of the second type of ordering, the one with positive vector chirality on the $J_2$ triangles, proceeds along the same lines. The results are summarized in Tables~\ref{tab:achiral} and \ref{tab:chiral}. The main difference is that in the former case half of the nearest neighbor triangles have non-zero toroidicity, which however averages to zero, while in the latter the same is true for the second-neighbor trimers. 

Adding the ring exchange which we found to be sizeable does not alter this ground state. 
Indeed, it is easy to show that if an AF coupling on a triangle $J>7L/2$ the ground state is not altered by adding the ring exchange.
Finally, interaction along the $c$ axis is strongly frustrated, with comparable $J_0'$, $J_1'$ and $J_1''$, which can lead, generally speaking, to spiral states propagating in this direction (given the higher coordination number for the last two). 

While these two states have been discussed above in terms of supercells, one can notice that they also form spin spirals of a sort. Namely, the first one can be described, using our notations, as the $(120,0,0)[4\pi/3,4\pi/3, 0]$ spiral, and the second as the $(0,120,120)[2\pi/3,2\pi/3,0]$ one. Note that
in this Hamiltonian, the two spirals are degenerate, and have not been included in our previous calculations and fitting. Thus, these two states are true predictions and can be explicitly verified. Indeed, we found that, as predicted 
by the model Hamiltonian, they are (a) degenerate within computational accuracy and (b) 25.6\,meV below the lowest-energy state found in the original calculations 
(namely, the standard 120$^\circ$ structure 
alternating antiferromagnetically between the planes.)

The last observation relates to the situation when the ordering planes are not parallel, $\mathbf{\Omega}\nparallel\boldsymbol{\omega}$, a finite local scalar chirality can be acquired, leading, for instance, to a topological Hall effect. This opens the door to a fluctuation-induced topological Hall effect at finite temperature\cite{10.1126/sciadv.abe2680}; however, further analysis is outside the scope of this paper.

\begin{figure*}[htbp]
	\centering
	\includegraphics[width=0.9\textwidth]{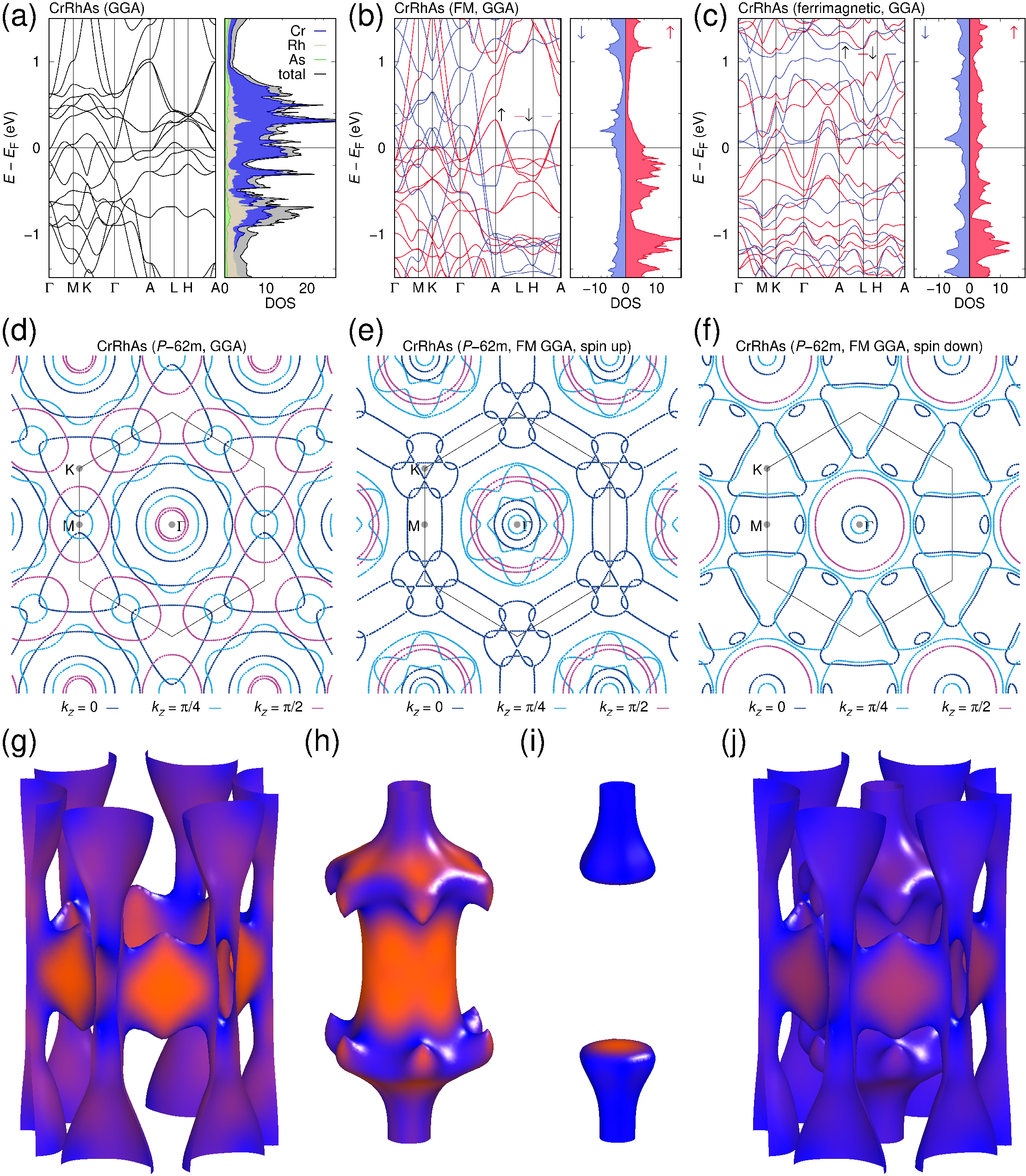}
	\caption{GGA electronic structure of CrRhAs. (a) Non-magnetic, (b) ferromagnetic and (c) ferrimagnetic bands with corresponding densities of states. (d) $k_x-k_y$ plane cuts of the Fermi surface for three different values of $k_z$. (e), (f) the same for the ferromagnetic solution. (g)-(j) 3D contour plots of the non-magnetic Fermi surface; (g)-(i) are three individual FS sheets, and (j) shows all of them together. In (g)-(j), color indicates Fermi velocity where blue is low, orange high.}
	\label{fig:CrRhAsBandCompose}
\end{figure*}

\section{Comparison with experiment}
Experimental information on this material is, basically, limited to four papers from T. Kaneko and co-authors~\cite{Ohta1990,Kanomata1991,Kaneko1992,Ohta1995}. It has been established that CrRhAs experiences an antiferromagnetic transition, with the N\'eel temperature reported at $T_{\rm N}=165$\,K \cite{Ohta1990,Kaneko1992,Kanomata1991} or 172\,K \cite{Ohta1995}. The antiferromagnetic order has not been established. Interestingly, the magnetic susceptibility measured in an interval between $T_{\rm N}$ and room temperature is distinctly non-Curie-Weiss (CW). In the first publication, Ref. \cite{Ohta1990}, it was fitted to the CW law, $\chi=C/(T-\Theta)+const$, but with a background of 1.47$\times 10^{-3}$ emu/mole, which, if interpreted as Pauli susceptibility, corresponds to 45 states/eV$\cdot$formula, or to an {\it ad hoc} formula $\chi=C'/(T-\Theta')^\gamma$, with $C' = 4.8\times 10^{-3}$ emu K/mole, $\Theta' = 20$ K and $\gamma = 0.16$. In a later paper, Ref. \cite{Kanomata1991}, the same formula was used with $\chi=C'/(T-\Theta')^\gamma$, with $C' = 3.2\times 10^{-2}$ emu K/mole, $\Theta' = -16$ K and $\gamma = 0.44$, presumably, due to a different protocol for the background removal. Either way, $\chi^{-1}(T)$ is strongly nonlinear and its slope gets smaller with the temperature. If one defines ``instantaneous'' CW parameters as $C(T)=1/(d\chi^{-1}(T)/dT)$ and $\Theta(T)=T-C(T)/\chi(T)$, then $\Theta(T)$ is becoming increasingly antiferromagnetic with temperature, and $C(T)$ also grows, corresponding to increasingly large effective moments.

All these observation, as strange as they may seem on the first glance, find natural explanations in our Hamiltonian and proposed ground state. Indeed, the former is dominated by the very strong $J_2$ interaction, which itself corresponds to a temperature scale of $2J_2$ (2 for the coordination number) of the order of 1000\,K, or if converted to the CW temperature and assuming spin 3/2 and the quantum factor $(S+1)/S=5/3$, corresponds to $T_{CW}=-578$\,K. This indicates that at room temperature the isolated trimers formed by the $J_2$ bonds are still strongly correlated, forming complexes with strongly suppressed net magnetic moment. As a result, the true CW regime is not attained until $T\agt 600$\,K, and the observed behavior is nothing but a graduate crossover from the fully correlated trimers with the effective moment $m_{\rm eff}\ll 1\ \mu_{\rm B}$ and $T_{\rm CW}$ defined by the other (besides $J_2$) interactions in the system (which is on the order of $-72$\,K), and the very high temperature regime, not reached in the reported experiments, where $m_{\rm eff}\sim \sqrt{3\cdot 5}=3.87$ $\mu_B$, and $T_{\rm CW}\sim -650$\,K. Finally, the relatively small value of $T_{\rm N}$ compared to the high $T_{\rm CW}$ temperature finds a natural explanation in the fact that the intra-trimer ordering that does happen at high temperature is not related to the temperature at which the individual triangles order with respect to each other; the latter is determined by the much weaker inter-trimer interactions. In fact, an upper boundary on the mean-field transition temperature can be derived by taking all interactions but $J_2$ with the same sign (remember that, for instance, $J_1$ and $J_3$, albeit being of the opposite signs, cooperate in the suggested ordering). This gives an estimate for the maximally possible ordering temperature of 346 K. The experimental number is right between the lower estimate of 72 K and this upper bound.

\section{Electronic structure of C\lowercase{r}R\lowercase{h}A\lowercase{s}}

Finally, we turn our attention to the band structure and Fermi surface of CrRhAs. We use GGA calculations with the FPLO basis to determine both non-magnetic (Fig.~\ref{fig:CrRhAsBandCompose}\,(a)) and examples of magnetic band structures (Fig.~\ref{fig:CrRhAsBandCompose}\,(b)-(c)). At the Fermi level, most of the density of states derives from Cr $3d$, with only small Rh $4d$ and very small As $3p$ contributions (Fig.~\ref{fig:CrRhAsBandCompose}\,(a)). In the nonmagnetic bands, a Dirac point at the K point can be seen about 0.4\,eV above the Fermi level but flat bands are hard to make out. CrRhAs remains metallic in both ferromagnetic and ferrimagnetic states (Fig.~\ref{fig:CrRhAsBandCompose}\,(b)-(c)) as well as in antiferromagnetic spin configurations (not shown). However, DOS at $E_{\rm F}$ is substantially lower in magnetic compared to nonmagnetic states. The nonmagnetic Fermi surface (see Fig.~\ref{fig:CrRhAsBandCompose}\,(d) for cuts, Fig.~\ref{fig:CrRhAsBandCompose}\,(h)-(i) for 3D plots) has some cylinder-like 2D features but also significant variation along $k_z$. The FS for the ferromagnetic solution (Fig.~\ref{fig:CrRhAsBandCompose}\,(e)-(f)) is not much simpler.

\section{Conclusions}

We studied one member of a large family of $XYZ$ compounds with spacegroup $P\bar{6}2m$ that contains twisted kagome and trimerized (distorted triangular) lattices. As many as 70 of them have significant magnetism on a kagome lattice, and despite the distorted geometry, the magnetic interaction Hamiltonian remains the same as ideal kagome at the nearest neighbor level, which makes this series of compounds a fertile playground for kagome physics. We used CrRhAs as an example to study different spin spiral and noncollinear energies.

To this end, we have calculated within the density functional theory the total energies of six different in-plane spin spirals, three different out-of-plane spirals, and two continuously varying non-collinear magnetic arrangements with the ${\bf q}=0 $ periodicity, a total of more than 230 first principle calculations. Based on these data, we generated a magnetic Hamiltonian that fits all these energies reasonably well (with max deviation within 15 meV). Interestingly, the resulting Hamiltonian was rather unusual in several aspects: first, we found that Heisenberg exchange interactions could not provide a satisfactory fit; adding ring exchange terms proved indispensable, especially for the two ${\bf q}=0 $ sets of calculations. Second, we found that the nearest neighbor exchange coupling was ferromagnetic, and thus not frustrated, but the leading (by far) interaction was the next nearest neighbor antiferromagnetic coupling in the twisted kagome plane, which is frustrated and leads to a curious, and, to the best of our knowledge, never discussed before magnetic Hamiltonian. The ground state of this Hamiltonian is controlled by independent triangles of a certain vector spin chirality, either positive or negative, but the same for all triangles, and varying toroidicity  between these triangles. Furthermore, it has a potential to develop, either statically due to spin-orbit coupling, or dynamically through topological field fluctuations, scalar spin chirality and topological Hall effect. These possibilities, however, go beyond the scope of our paper.

We hope that this study would motivate further experimental and theoretical research into this intriguing family, and particularly this specific system.

\acknowledgments{
Y.N.H. is supported by the National Natural Science Foundation of China (under Grant No. 11904319), and she thanks Bao Weicheng for valuable help in this research. I.I.M. acknowledges support from the U.S. Department of Energy through the grant No. DE-SC0021089.
}

\bibliography{XYZ_kagome}

\clearpage


\section*{Supplementary material}

# CrRhAs: a member of a large family of metallic kagome antiferromagnets
– Supplemental Material –

Y. N. Huang
*Department of Physics, Zhejiang University of Science and Technology, Hangzhou 310023, People's Republic of China*

Harald O. Jeschke
*Research Institute for Interdisciplinary Science, Okayama University, Okayama 700-8530, Japan*

Igor I. Mazin[*]
*Department of Physics & Astronomy, George Mason University, Fairfax, VA 22030, USA and
Quantum Science and Engineering Center, George Mason University, Fairfax, VA 22030, USA.*
(Dated: November 9, 2022)

## I. $XYZ$ COMPOUNDS WITH SIGNIFICANT MAGNETISM

Initially, we classified $XYZ$ compounds with space group $P\bar{6}2m$ (no. 189) by groups of elements. By analyzing all $P\bar{6}2m$ $XYZ$ compounds in the Materials Project database, we found that $X$, $Y$, $Z$ atoms belong to finite groups of elements as shown in Table S1.

With poor metal, we refer to Al, Ga, In, Sn, Tl, Pb, Bi. As metalloid, we classify B, Si, Ge, As and Sb. Relevant nonmetals are H, C, N and P. Specifically, the $XY$ combination of element series is shown in Table S2.

TABLE S1: $XYZ$ compounds classified by periodic table series

| Atom | Element Series |
|---|---|
| $X$ | actinide, alkaline earth metal, lanthanide, poor metal, transition metal |
| $Y$ | alkali metal, alkaline earth metal, lanthanide, metalloid, poor metal, transition metal |
| $Z$ | alkaline earth metal, lanthanide, metalloid, nonmetal, poor metal, transition metal |

TABLE S2: $XY$ combination of groups of elements.

| $X$ group | $Y$ group |
|---|---|
| actinide | metalloid |
| actinide | poor metal |
| alkaline earth metal | alkali metal |
| alkaline earth metal | poor metal |
| alkaline earth metal | transition metal |
| lanthanide | alkali metal |
| lanthanide | alkaline earth metal |
| lanthanide | metalloid |
| lanthanide | poor metal |
| lanthanide | transition metal |
| poor metal | lanthanide |
| poor metal | transition metal |
| transition metal | alkali metal |
| transition metal | alkaline earth metal |
| transition metal | metalloid |
| transition metal | poor metal |
| transition metal | transition metal |

We choose $P\bar{6}2m$ $XYZ$ compounds with significant magnetism from the Materials Project (MP)[S1], and organize them into Table S3. mp-id is the material ID of MP. $X$ is the kagome atom, $Y$ is trimer atom. $|v|(X)$ and $|v|(Y)|$ are

---

[*] imazin2@gmu.edu

the $v$ values of $X$ and $Y$ respectively where $v = x - 1/2$ is is determined from the fractional coordinate $x$ of the $X$ or $Y$ atom (see also main text). The ICSD column shows whether the compound exists and has been reported to the Inorganic Crystal Structure Database (ICSD).

TABLE S3: XYZ (space group $P\bar{6}2m$) compounds with significant magnetism.

| mp-id | compound | X | $|v|(X)$ | Y | $|v|(Y)|$ | ICSD |
|---|---|---|---|---|---|---|
| mp-20665 | CeInCu | Ce | 0.081851 | In | 0.252775 | yes |
| mp-15683 | CrRhAs | Cr | 0.092491 | Rh | 0.245181 | yes |
| mp-1205827 | CrPdAs | Cr | 0.103397 | Pd | 0.232872 | no |
| mp-1087537 | CrPdP | Cr | 0.105917 | Pd | 0.231023 | yes |
| mp-4989 | CrNiAs | Cr | 0.095091 | Ni | 0.244877 | yes |
| mp-1206397 | CrCoAs | Cr | 0.086577 | Co | 0.252504 | no |
| mp-1206083 | CrFeAs | Cr | 0.087428 | Fe | 0.253454 | no |
| mp-1079966 | CrNiP | Cr | 0.091987 | Ni | 0.244212 | yes |
| mp-1091389 | EuInPd | Eu | 0.089105 | In | 0.252248 | yes |
| mp-1225012 | FeRhAs | Fe | 0.103656 | Rh | 0.236774 | no |
| mp-1224852 | FeRhP | Fe | 0.110326 | Rh | 0.231339 | no |
| mp-1080653 | FeNiP | Fe | 0.098632 | Ni | 0.242936 | yes |
| mp-1080612 | GdAlCu | Gd | 0.085396 | Al | 0.267555 | yes |
| mp-20332 | GdMgPd | Gd | 0.086926 | Mg | 0.257178 | yes |
| mp-22028 | GdMgPt | Gd | 0.090582 | Mg | 0.25546 | yes |
| mp-623988 | GdCdPd | Gd | 0.093488 | Cd | 0.245842 | yes |
| mp-20224 | GdInPd | Gd | 0.090917 | In | 0.242642 | yes |
| mp-1101090 | GdAlNi | Gd | 0.080304 | Al | 0.267974 | yes |
| mp-1080178 | GdZnPd | Gd | 0.096443 | Zn | 0.249717 | yes |
| mp-30692 | GdTlPd | Gd | 0.09759 | Tl | 0.234001 | yes |
| mp-574004 | GdMgGa | Gd | 0.075155 | Mg | 0.256293 | yes |
| mp-20058 | GdAlPd | Gd | 0.082467 | Al | 0.265203 | yes |
| mp-1101056 | GdMgIn | Gd | 0.065156 | Mg | 0.259364 | yes |
| mp-1079690 | GdAgSi | Gd | 0.083068 | Ag | 0.249535 | yes |
| mp-1080496 | GdInAu | Gd | 0.091863 | In | 0.242823 | yes |
| mp-1079739 | GdSnPt | Gd | 0.089816 | Sn | 0.244624 | yes |
| mp-9341 | GdAgGe | Gd | 0.084516 | Ag | 0.249724 | yes |
| mp-19745 | GdInIr | Gd | 0.092568 | In | 0.245165 | yes |
| mp-623043 | GdInRh | Gd | 0.091525 | In | 0.245585 | yes |
| mp-1206951 | GdMgTl | Gd | 0.070661 | Mg | 0.256942 | no |
| mp-14210 | GdLiGe | Gd | 0.077382 | Li | 0.267255 | no |
| mp-582029 | GdInPt | Gd | 0.092445 | In | 0.249518 | yes |
| mp-580154 | GdMgAu | Gd | 0.08771 | Mg | 0.254229 | yes |
| mp-21458 | GdCdCu | Gd | 0.088142 | Cd | 0.250108 | yes |
| mp-1079035 | MnPdAs | Mn | 0.092059 | Pd | 0.238231 | yes |
| mp-1079184 | MnRhGe | Mn | 0.0912 | Rh | 0.241729 | yes |
| mp-10049 | MnRuAs | Mn | 0.095843 | Ru | 0.242341 | yes |
| mp-1206331 | MnPdP | Mn | 0.095237 | Pd | 0.235672 | no |
| mp-610972 | MnPdGe | Mn | 0.096695 | Pd | 0.235066 | yes |
| mp-567856 | MnRhAs | Mn | 0.095212 | Rh | 0.236511 | yes |
| mp-1079845 | MnNiAs | Mn | 0.085264 | Ni | 0.248649 | yes |
| mp-1079806 | MnTiAs | Mn | 0.109278 | Ti | 0.238304 | yes |
| mp-1079405 | MnRhP | Mn | 0.10521 | Rh | 0.235009 | yes |
| mp-4238 | MnFeAs | Mn | 0.094506 | Fe | 0.245038 | yes |
| mp-975423 | MnNiP | Mn | 0.100394 | Ni | 0.24237 | yes |
| mp-20314 | NpSnIr | Np | 0.085981 | Sn | 0.245747 | yes |
| mp-1091393 | PuGaNi | Pu | 0.082098 | Ga | 0.260442 | yes |
| mp-1078774 | PuGaRh | Pu | 0.086028 | Ga | 0.252119 | yes |
| mp-1079793 | PuAlCo | Pu | 0.071542 | Al | 0.262264 | yes |
| mp-1080562 | USnPt | U | 0.087948 | Sn | 0.245606 | yes |
| mp-1078683 | UInPd | U | 0.080172 | In | 0.251386 | yes |
| mp-1078582 | UInPt | U | 0.08236 | In | 0.249319 | yes |
| mp-1080442 | UGaPd | U | 0.073034 | Ga | 0.261463 | yes |
| mp-1079696 | USnRh | U | 0.081128 | Sn | 0.248363 | yes |
| mp-1080803 | USnIr | U | 0.083744 | Sn | 0.247099 | yes |
| mp-1095127 | USbRu | U | 0.086204 | Sb | 0.242608 | yes |

TABLE S3: XYZ (space group P$\bar{6}2m$) compounds with significant magnetism.

| | | | | | | |
|---|---|---|---|---|---|---|
| mp-1091404 | UInRh | U | 0.081975 | In | 0.253257 | yes |
| mp-21494 | USnCo | U | 0.081587 | Sn | 0.249041 | yes |
| mp-1080077 | UGaPt | U | 0.078517 | Ga | 0.256914 | yes |
| mp-1079949 | UAlPt | U | 0.074677 | Al | 0.262571 | yes |
| mp-5015 | UAlRh | U | 0.074114 | Al | 0.265481 | yes |
| mp-1078870 | UGaRh | U | 0.078603 | Ga | 0.260529 | yes |
| mp-1079086 | UAlNi | U | 0.071447 | Al | 0.264623 | yes |
| mp-21320 | UGaNi | U | 0.071759 | Ga | 0.261885 | yes |
| mp-1079219 | UAlIr | U | 0.07841 | Al | 0.263225 | yes |
| mp-1080013 | UAlRu | U | 0.074845 | Al | 0.266171 | yes |
| mp-1078880 | UGaRu | U | 0.076241 | Ga | 0.262093 | yes |
| mp-19811 | USnRu | U | 0.083475 | Sn | 0.248264 | yes |
| mp-1078586 | UGaCo | U | 0.0795095 | Ga | 0.2635175 | yes |

## II. NEAREST NEIGHBOR INFORMATION OF CRRHAS

We summarize the nearest neighbor information to a reference Cr1 atom of CrRhAs in Table S4. The in-plane distance is the projected bond length in the $xy$ plane. $n_\text{bonds}$ is the number of bonds that satisfy the in-plane distance constraint. $d_z = 0\,\text{Å}$, $d_z = 3.718\,\text{Å}$, $d_z = 7.436\,\text{Å}$, $d_z = 11.154\,\text{Å}$ correspond to successive Cr plane distances along the $z$ direction, and $Ji$ in each column is the hopping from the reference Cr1 atom to the corresponding Cr plane.

TABLE S4. Nearest neighbor information of CrRhAs. $d_z$ are distances between Cr planes along $z$ direction. Distance $d_\text{Cr-Cr}$ is given in brackets for easy identification of the bond. In the naming scheme of the main paper, the first column of couplings is called $J_0, J_1, J_2, \ldots$, the second column is called $J'_0, J'_1, J'_2, \ldots$, the third $J''_0, J''_1, J''_2, \ldots$, and so on.

| $d_\text{inplane}$ (Å) | Cr1 bond to | $n_\text{bonds}$ | $d_z = 0\,\text{Å}$ | $d_z = 3.718\,\text{Å}$ | $d_z = 7.436\,\text{Å}$ | $d_z = 11.154\,\text{Å}$ |
|---|---|---|---|---|---|---|
| 0. | {Cr1} | 1 | J0 (0 Å) | J2 (3.718 Å) | J9 (7.436 Å) | J25 (11.154 Å) |
| 3.384 | {Cr3,Cr2} | 4 | J1 (3.384 Å) | J4 (5.028 Å) | J12 (8.170 Å) | J26 (11.656 Å) |
| 4.404 | {Cr3,Cr2} | 2 | J3 (4.404 Å) | J5 (5.764 Å) | J14 (8.642 Å) | J29 (11.992 Å) |
| 6.384 | {Cr1} | 6 | J6 (6.384 Å) | J8 (7.388 Å) | J17 (9.800 Å) | J34 (12.852 Å) |
| 6.653 | {Cr2,Cr3} | 2 | J7 (6.653 Å) | J10 (7.622 Å) | J19 (9.978 Å) | J35 (12.988 Å) |
| 7.756 | {Cr3,Cr2} | 4 | J11 (7.756 Å) | J13 (8.601 Å) | J22 (10.745 Å) | J40 (13.585 Å) |
| 9.221 | {Cr3,Cr2} | 4 | J15 (9.221 Å) | J18 (9.942 Å) | J28 (11.846 Å) | J45 (14.472 Å) |
| 9.642 | {Cr3,Cr2} | 4 | J16 (9.642 Å) | J20 (10.334 Å) | J30 (12.176 Å) | J47 (14.744 Å) |
| 10.433 | {Cr3,Cr2} | 4 | J21 (10.433 Å) | J24 (11.076 Å) | J33 (12.812 Å) | J52 (15.273 Å) |
| 11.057 | {Cr1} | 6 | J23 (11.057 Å) | J27 (11.666 Å) | J38 (13.325 Å) | J58 (15.706 Å) |
| 12.593 | {Cr2,Cr3} | 4 | J31 (12.593 Å) | J36 (13.131 Å) | J46 (14.625 Å) | J68 (16.823 Å) |
| 12.768 | {Cr1} | 6 | J32 (12.768 Å) | J37 (13.298 Å) | J48 (14.776 Å) | J70 (16.954 Å) |

## III. HAMILTONIAN COEFFICIENT EXPRESSIONS FOR $(q_x, 0, q_z)$ SPIRALS

We derived coefficient expressions $C_i, C'_i, C''_i, ...$ and $D, D', D''$ for Eq. (4) in the main text.

TABLE S5. Coefficient expression of $C_i$ and $D$ for $(q_x, 0, q_z)$ spin spirals.

| | |
|---|---|
| $E_0$ | $1$ |
| $C_1$ | $\cos(\theta_x - \phi(1) + \phi(2)) + \cos(\theta_x + \phi(2) - \phi(3)) + \cos(\phi(1) - \phi(2)) + 2\cos(\phi(1) - \phi(3)) + \cos(\phi(2) - \phi(3))$ |
| $C_2$ | $\cos(\theta_x - \phi(1) + \phi(2)) + \cos(\theta_x - \phi(1) + \phi(3)) + \cos(\phi(2) - \phi(3))$ |
| $C_3$ | $3(2\cos(\theta_x) + 1)$ |
| $C_4$ | $\cos(\theta_x + \phi(1) - \phi(3)) + \cos(\theta_x + \phi(2) - \phi(3)) + \cos(\phi(1) - \phi(2))$ |
| $C'_0$ | $3\cos(\theta_z)$ |
| $C'_1$ | $2\cos(\theta_z)(\cos(\theta_x - \phi(1) + \phi(2)) + \cos(\theta_x + \phi(2) - \phi(3)) + \cos(\phi(1) - \phi(2)) + 2\cos(\phi(1) - \phi(3)) + \cos(\phi(2) - \phi(3)))$ |
| $C'_2$ | $2\cos(\theta_z)(\cos(\theta_x - \phi(1) + \phi(2)) + \cos(\theta_x - \phi(1) + \phi(3)) + \cos(\phi(2) - \phi(3)))$ |
| $C'_3$ | $6(2\cos(\theta_x) + 1)\cos(\theta_z)$ |
| $C'_4$ | $2\cos(\theta_z)(\cos(\theta_x + \phi(1) - \phi(3)) + \cos(\theta_x + \phi(2) - \phi(3)) + \cos(\phi(1) - \phi(2)))$ |
| $C''_0$ | $3\cos(2\theta_z)$ |
| $C''_1$ | $2\cos(2\theta_z)(\cos(\theta_x - \phi(1) + \phi(2)) + \cos(\theta_x + \phi(2) - \phi(3)) + \cos(\phi(1) - \phi(2)) + 2\cos(\phi(1) - \phi(3)) + \cos(\phi(2) - \phi(3)))$ |
| $C''_2$ | $2\cos(2\theta_z)(\cos(\theta_x - \phi(1) + \phi(2)) + \cos(\theta_x - \phi(1) + \phi(3)) + \cos(\phi(2) - \phi(3)))$ |
| $C''_3$ | $6(2\cos(\theta_x) + 1)\cos(2\theta_z)$ |
| $C''_4$ | $2\cos(2\theta_z)(\cos(\theta_x + \phi(1) - \phi(3)) + \cos(\theta_x + \phi(2) - \phi(3)) + \cos(\phi(1) - \phi(2)))$ |
| $D_1$ | $\cos(\phi(1) - \phi(3))\cos(\theta_x - \phi(1) + \phi(2)) + \cos(\phi(1) - \phi(3))(\cos(\theta_x + \phi(2) - \phi(3)) + \cos(\phi(2) - \phi(3)) + 1) + \cos(\theta_x)\cos(\theta_x - \phi(1) + 2\phi(2) - \phi(3)) + \cos(\phi(1) - \phi(2))\cos(\phi(1) - \phi(3))$ |
| $D_2$ | $\frac{1}{2}(\cos(\phi(2) - \phi(3))(2\cos(\theta_x - \phi(1) + \phi(2)) + 2\cos(\theta_x - \phi(1) + \phi(3)) + 1) + \cos(2\theta_x - 2\phi(1) + \phi(2) + \phi(3)))$ |

The definition of $\phi(1), \phi(2), \phi(3)$ and $\theta_x$ is illustrated in Fig. S1. For $\theta_z$, it is a common rotation angle between current plane and the next plane of all moments.

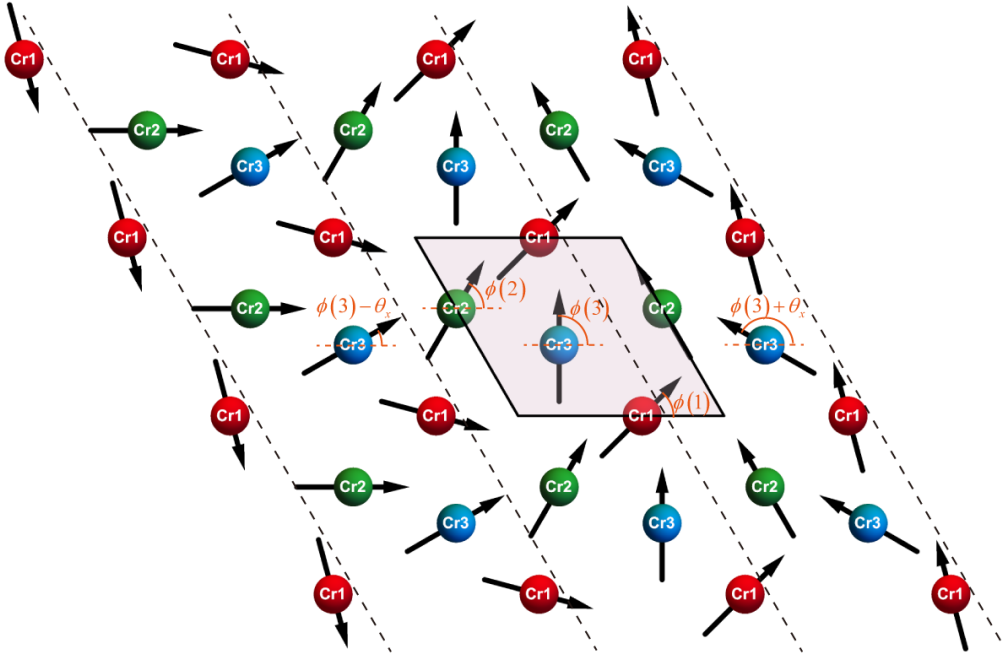

FIG. S1. An illustration of a $(q_x, 0, 0)$ spiral. $\phi(1), \phi(2), \phi(3)$ are the polar angles of Cr1, Cr2 and Cr3 moments. $\theta_x$ is the spiral angle going from one strip marked by dashed lines to the next one.

## IV. ADDITIONAL ATTEMPTS OF FITTING DFT TOTAL ENERGIES

Figure S2 shows a simple $\cos(\alpha)$ fit of the noncollinear periodic calculations (the $++\alpha$ and $+-\alpha$ series). The deviations of the fit indicate that higher harmonics are necessary for a good fit, *i.e.* a pure Heisenberg Hamiltonian cannot completely represent the noncollinear DFT energies.

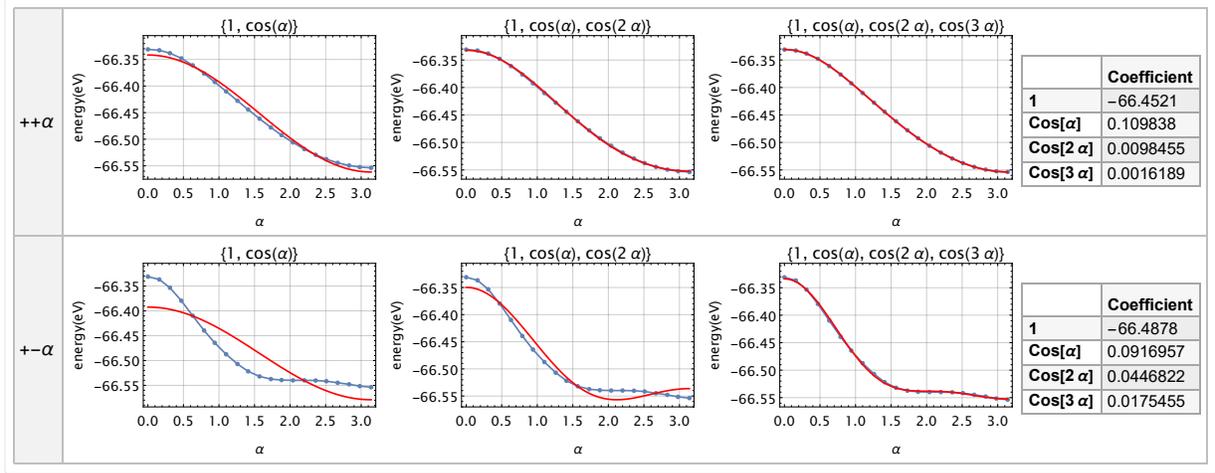

FIG. S2. Pure cosine fitting for $++\alpha$ and $+-\alpha$ series

In Figure S3, we show a fit of all spiral energies with Heisenberg exchange but without ring exchange terms. The quality of fit is clearly lower than the full fit shown in Fig. 6 of the main text, especially at $\theta = 0$ for $(120)[q00]$ and $(120)[00q]$.

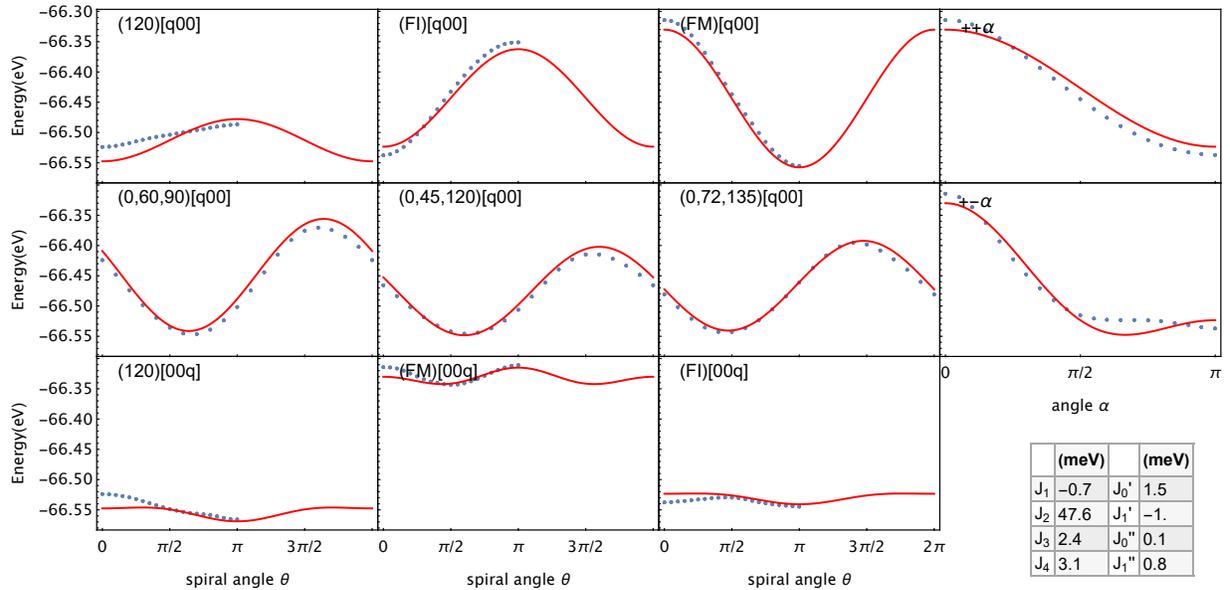

FIG. S3. Fitting all spiral energy curves without ring exchange



\end{document}